\documentclass[%
superscriptaddress,
twocolumn,
amsmath,amssymb,
 aps,
prb,
floatfix,
]{revtex4-2}

\usepackage{siunitx}
\usepackage{graphicx}
\usepackage{multirow}
\usepackage{amsmath}
\usepackage{booktabs}
\usepackage[colorlinks]{hyperref}
\hypersetup{%
	plainpages=true,
	breaklinks=true,
	hypertexnames=false,
	pageanchor=true,
	colorlinks=true,
	linkcolor={blue},
	citecolor={red},
	urlcolor={blue},
	anchorcolor={black}
}

\usepackage[dvipsnames]{xcolor}
\usepackage[normalem]{ulem}

\mathchardef\mhyphen="2D

\DeclareMathOperator{\MLP}{MLP}
\DeclareMathOperator{\logcosh}{log-cosh}
\DeclareMathOperator{\Tr}{Tr}
\renewcommand{\Re}{\operatorname{Re}}
\renewcommand{\Im}{\operatorname{Im}}
\newcommand{\software}[1]{\textsc{#1}}

\begin{document}
\title{
Transformer Wave Function for Quantum Long-Range models
}

\author{Sebastián Roca-Jerat}
\affiliation {Instituto de Nanociencia y Materiales de Aragón (INMA), CSIC-Universidad de Zaragoza, Zaragoza 50009, Spain}
\affiliation{Departamento de Física de la Materia Condensada, Universidad de Zaragoza, Zaragoza 50009, Spain}

\author{Manuel Gallego}
\affiliation {Instituto de Nanociencia y Materiales de Aragón (INMA), CSIC-Universidad de Zaragoza, Zaragoza 50009, Spain}
\affiliation{Departamento de Física Teórica, Universidad de Zaragoza, Zaragoza 50009, Spain}

\author{Fernando Luis}
\affiliation {Instituto de Nanociencia y Materiales de Aragón (INMA), CSIC-Universidad de Zaragoza, Zaragoza 50009, Spain}
\affiliation{Departamento de Física de la Materia Condensada, Universidad de Zaragoza, Zaragoza 50009, Spain}

\author{Jesús Carrete}
\affiliation {Instituto de Nanociencia y Materiales de Aragón (INMA), CSIC-Universidad de Zaragoza, Zaragoza 50009, Spain}
\affiliation{Departamento de Física de la Materia Condensada, Universidad de Zaragoza, Zaragoza 50009, Spain}

\author{David Zueco}
\email{dzueco@unizar.es}
\affiliation {Instituto de Nanociencia y Materiales de Aragón (INMA), CSIC-Universidad de Zaragoza, Zaragoza 50009, Spain}
\affiliation{Departamento de Física de la Materia Condensada, Universidad de Zaragoza, Zaragoza 50009, Spain}
\date{\today}

\begin{abstract}
We employ a neural-network architecture based on the Vision Transformer (ViT) architecture to find the ground states of quantum long-range models, specifically the transverse-field Ising model for spin-1/2 chains across different interaction regimes. Harnessing the transformer's capacity to capture long-range correlations, we compute the full phase diagram and critical properties of the model, in both the ferromagnetic and antiferromagnetic cases. Our findings show that the ViT maintains high accuracy across the full phase diagram. We compare these results with previous numerical studies in the literature and, in particular, show that the ViT has a superior performance than a restricted-Boltzmann-machine-like ansatz.
\end{abstract}

\maketitle

\section{Introduction}
Understanding quantum many-body problems is crucial for elucidating the properties of condensed matter systems. However, this is generally a challenging task, as exemplified by the problem of searching for the ground states of strongly correlated models. Beyond exactly solvable models, some systems are tractable by introducing elements  such as matrix product states (MPS), tensor networks \cite{Orus2014}, dynamical mean field theory (DMFT) \cite{Georges1996}, quantum Monte Carlo \cite{Becca2017}, and hybrid quantum-classical algorithms \cite{alexeev2023quantumcentric}. Additionally, there are physically inspired ansätze available, like the Bethe and polaron approaches, correlator product states \cite{changlani2009}, and generalized coherent states \cite{anders2006}.
\\
In 2017, neural networks (NNs) emerged as an alternative ansatz for solving quantum many-body systems \cite{Carleo2017}. NNs have since been recognized for their ability to generate states beyond the area law entanglement \cite{Deng2017, Glasser2018}, addressing problems beyond the reach of MPS. Variational optimization of ansätze like NNs also avoids the sign problem of other quantum Monte Carlo techniques, enabling the handling of frustrated systems and account for states with long correlation lengths beyond those tractable with DMFT. Despite these advantages, it is difficult to determine in advance which NN architecture will be most suitable for a specific problem or whether it can be trained with sufficient precision. Therefore, it is interesting to investigate generic models of significant importance and evaluate how well they are solved using NNs \cite{medvidovic2024}.

In this context, we focus on a set of models in which two-body interactions decay according to a power law, commonly referred to as long-range systems. They are ubiquitous in nature, with examples arising, among others, from dipolar \cite{cartarius2014structural}, Coulomb \cite{abergel2009longrange}, and van der Waals interactions \cite{beguin2013direct}. Recent experimental advances in atomic, molecular, and optical systems have renewed interest in long-range models \cite{Britton2012, knap2013probing, monroe2021programmable, browaeys2020manybody}. In these experiments, the effective interactions are often long ranged and tunable, underscoring the need for a comprehensive understanding of long-range systems.
Although less studied than their short-ranged counterparts, there are already some rigorous and numerical results available for systems with long-range interactions \cite{mukamel2008statistical, campa2009statistical, fey2016critical, defenu2023longrange, defenu2023outofequilibrium}. Various equilibrium and dynamical properties have been explored in comparison with short-ranged systems. Notable examples include the existence (or absence) of an area law of entanglement \cite{koffel2012entanglement, kuwahara2020area, vodola2014kitaev, ares2018entanglement}, the algebraic decay of two-point correlators outside criticality \cite{vodola2015longrange, vanderstraeten2018quasiparticles, francica2022correlations}, the spreading of correlations \cite{schneider2022spreading}, the existence of Majorana modes \cite{jager2020edge}, and topological properties \cite{viyuela2016topological}. 

In this work, we test neural networks (NNs) for solving quantum long-range models, specifically focusing on the quantum Ising model as a paradigmatic example. Our primary motivation is to use a sufficiently generic NN ansatz capable of covering the entire phase diagram, from short-range to strong long-range interactions, and for both ferromagnetic and antiferromagnetic phases.
Since the introduction of Restricted Boltzmann Machines (RBMs), several NN architectures have been developed, including multilayer perceptrons \cite{Saito2017}, convolutional neural networks \cite{choo2019}, and recurrent architectures \cite{sharir2020deep}. This list is not exhaustive; for a comprehensive review of the field, we refer to Ref. \citenum{medvidovic2024}. Recently, the transformer \cite{Vaswani2017} has proven to be a game changer in the field of machine learning (ML), replacing recurrent architectures in natural language processing tasks \cite{raffel2020exploring}, being adapted with particular success adapted to computer vision tasks in the form of the Vision Transformer (ViT) \cite{vision_transformer} but also to more specialized applications like the determination of protein structures \cite{jumper2021highly}. A key feature of transformer architectures is their decoupling of nonlinear processing, which may still be performed by conventional fully connected subnetworks acting on subsets of the data, and mixing between those pieces of data according to the relevance of those interactions to the result, which is handled by an attention mechanism. This allows it to extract complex patterns very efficiently. Due to these promising features, application of transformers in the field of condensed matter as a variational ansatz has quickly followed \cite{Viteritti2023, Sprague2024, Luo2023}. This type of architecture could help finding solutions more efficiently in problems where established methods like MPS (systems with more than one dimension) or quantum Monte Carlo encounter difficulties (wave function with negative coefficients).

In this work, we continue to explore the capabilities of this type of architecture by selecting the Vision Transformer (ViT).
We demonstrate that this architecture is capable of computing the full phase diagram and characterizing its critical properties. By doing so, we provide a single ansatz that encompasses previous numerical works discussing parts of the phase diagram \cite{koziol2021quantumcritical, zhu2018fidelity, romanroche2023, sun2017, kim2023neural}. Additionally, we compare the performance of our ViT and RBM architectures. Our results help establish transformers as an efficient variational ansatz that can be highly useful for tackling a wide range of problems in many-body physics.

The rest of the manuscript is organized as follows: in section~\ref{sec:LRmodel} we lay out the details of the family of models we are trying to solve, in section~\ref{sec:methods} we discuss the methods, software and parameters that we use, in section~\ref{sec:results} we present and contextualize our results, and in the last section we summarize our main conclusions.
Details of the finite size scaling are sent to Appendix \ref{app:FSSA}.


\section{Long-range interacting quantum models}\label{sec:LRmodel}

\begin{figure}
    \centering
    \includegraphics[width = \columnwidth]{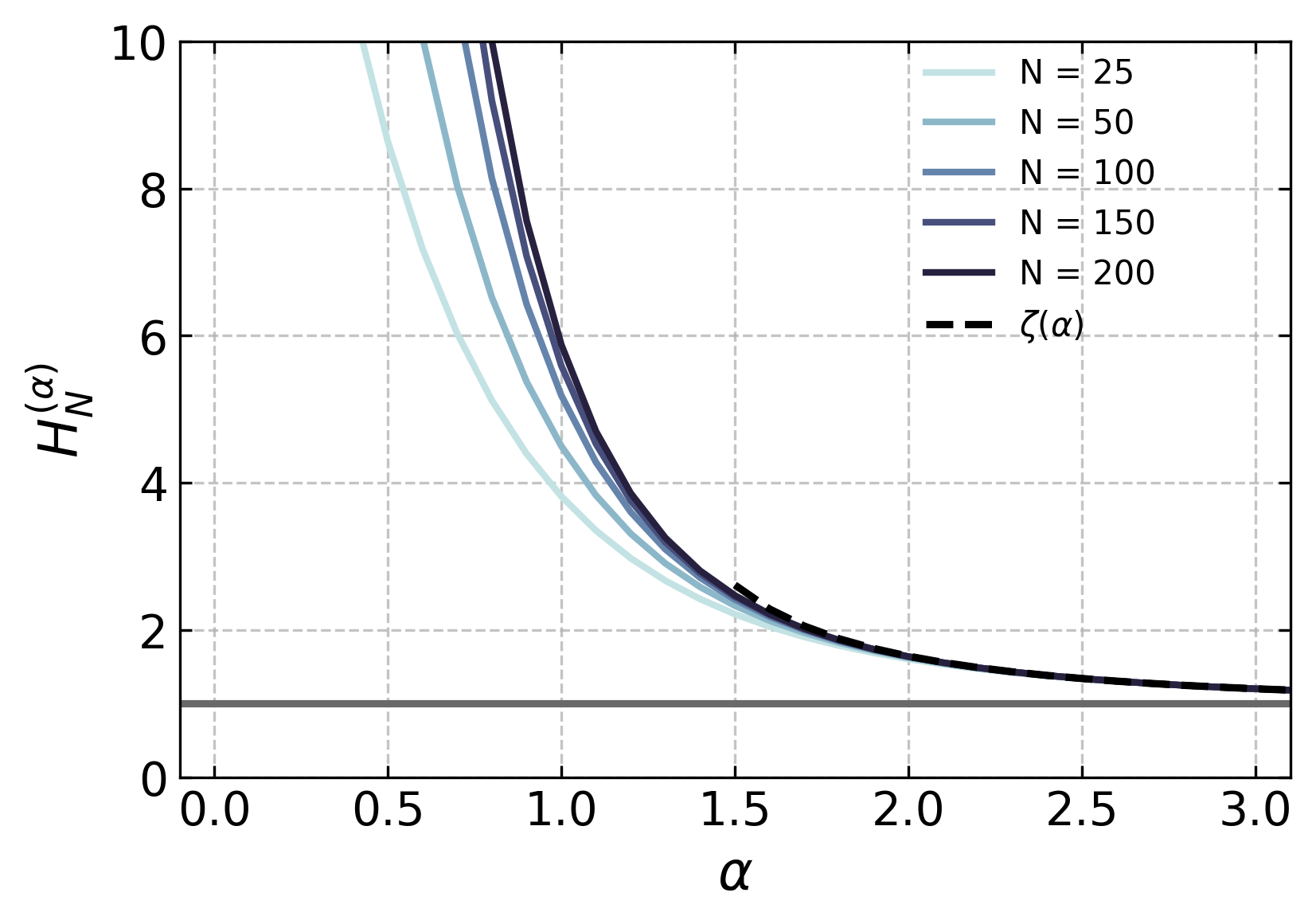}
    \caption{Dependence of the generalised harmonic numbers on the system's size ($N$) and on the range of the interactions  ($\alpha$). It can be seen that for $\alpha < 1.5$ the dependence of this factor on size intensifies. The solid grey line indicates the asymptotic value of $H_{N}^{(\alpha)} = 1$ and the dashed black line indicates the Riemann zeta function, $\zeta(\alpha)$, to which the generalised harmonic number tends when $N \to \infty$.}
    \label{fig:Hnalpha}
\end{figure}

Generically, quantum long-range models on a $d$-dimensional, $N$-site lattice contain interactions described by a Hamiltonian of the form
\begin{equation}
    \mathcal H_{\rm c}
    =
    -\frac{1}{2}\sum_{ij}^N J_{ij} \left( \mathcal C_i \mathcal C^\dagger_j + h.c.\right)\,,    \label{eq:longrangecoupling}
\end{equation}
where $\mathcal C_i$ is a local operator acting on site $i$ and the interactions decay according to a power law
\begin{equation}
\label{eq:Jalpha}
J_{ij} =  J \frac{ \tilde J_{ij}} {\tilde N},
\end{equation}
with
\begin{equation}
    \tilde J_{ij}=  \begin{cases}
    b & \text { if } \; i=j \\
    r_{ij}^{-\alpha} & \text { otherwise} .
    \end{cases}
\end{equation}
In this equation, $b$ is a parameter that can be tuned to shift the spectrum of $J$ and is taken as $b = 1$ in our simulations, and the parameter $\alpha$ sets the range of the interactions. Throughout this paper, we will assume periodic boundary conditions (PBC) and $d=1$, therefore we define $r_{ij} = \min\left(|i-j|, N - |i-j|\right)$.

It is important to discuss the normalization used in Eq.~\eqref{eq:Jalpha}. For $\alpha < d$, the interactions decay slowly enough that the sum in the coupling term \eqref{eq:longrangecoupling} depends on $N$ superlinearly, breaking the extensive character of the model \cite{campa2009statistical, campa2014physics, defenu2023longrange}. Kac's renormalization factor, $1/\tilde{N}$, restores that feature, ensuring a well-defined thermodynamic limit. In particular,
\begin{equation}\label{eq:Kac_factor}
\tilde N = \sum_j \tilde J_{ij}  = 1 + \sum_{j=1}^N \frac{1}{r_{ij}^\alpha} = 1+ H_{\lfloor N/2\rfloor}^{(\alpha)} + H_{N-1-\lfloor N/2\rfloor}^{(\alpha)} \; ,
\end{equation}
where $H_{N}^{(\alpha)}$ is a generalized harmonic number. We have used the fact PBC make the model translationally invariant and thus $\sum_i \tilde J_{ij}$ is independent of $j$. In figure \ref{fig:Hnalpha}, we plot $H_N^{(\alpha)}$ for different sizes as a function of $\alpha$. For $\alpha>1$, $H_{N}^{(\alpha)}$ approaches the $\zeta(\alpha)$ function as $N \to \infty$. Indeed, for $\alpha \gtrapprox 2$, in the next sections we confirm through numerical simulations that the dependence of the Kac factor on system size can be safely neglected. Eventually, $\tilde{N} \to 3$ as $\alpha \to \infty$. For $\alpha < 1$, the asymptotics of $H_{N}^{(\alpha)}$ are richer, but it diverges as a function of $N$, which makes this regularization necessary to have a well-defined thermodynamic limit. 

\begin{figure}
    \centering
    \includegraphics[width = 1.0 \columnwidth]{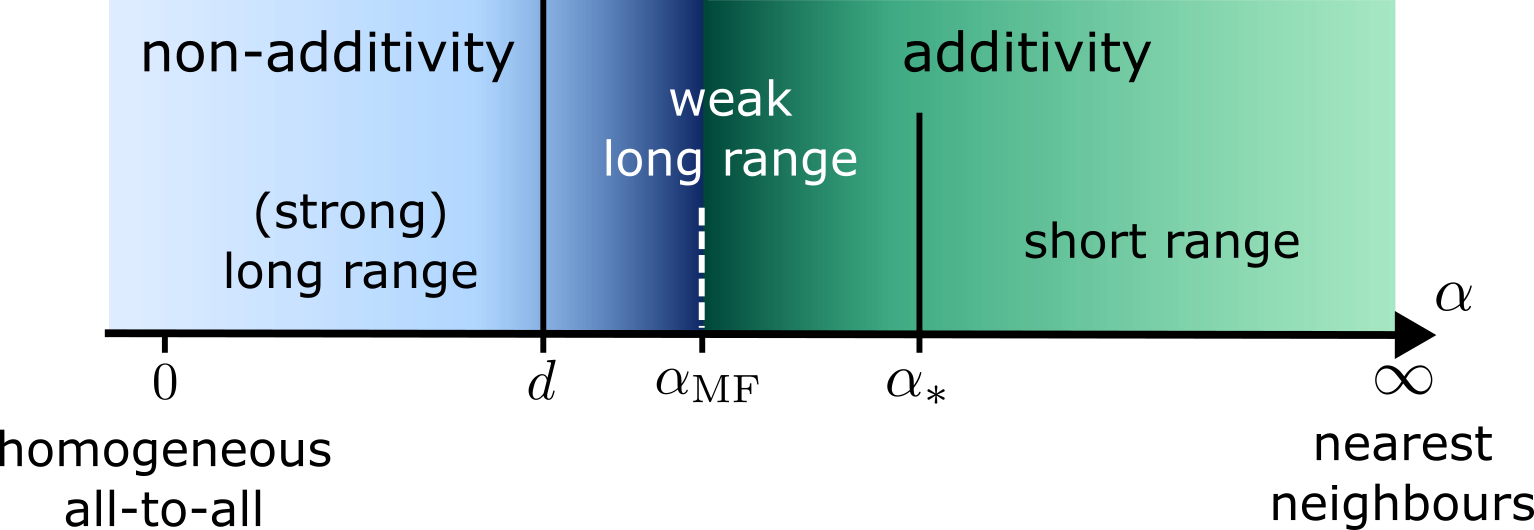}
    \caption{Diagram of the nature of interactions as a function of $\alpha$.}
    \label{fig:longcla}
\end{figure}

The general phenomenology, both quantum and classical, can be characterized by the value of $\alpha$.
The non-additive regime, characterized by $\alpha < 1$, is termed \emph{(strong) long-range}. This regime is often overlooked in many analytical and numerical  studies due to the challenges posed by an ill-defined thermodynamic limit resulting from non-extensivity. However, Kac's rescaling overcomes this obstacle.
In the quantum context,  numerical studies that solve the transverse-field Ising model in the strong long-range regime can be found in Refs.~\onlinecite{koziol2021quantumcritical, gonzalezlazo2021finitetemperature}, confirming that in this regime, the model is within the mean-field universality class. Furthermore, it has been proven that, in both the classical and quantum realms, some models can be solved exactly through mean-field-like approximations that become exact in this regime \cite{campa2000canonical,mori2011instability,mori2010analysis,mori2012microcanonical,mori2012equilibrium, romanroche2023}.
\\
Within $\alpha > 1$ domain, additional subregimes are discernible: a threshold $\alpha_*$ exists such that for $\alpha > \alpha_*$ the critical exponents of the model align with those of the nearest-neighbour model ($\alpha \to \infty$). This is the \emph{short-range regime}. Conversely, in the range where $1 < \alpha < \alpha_*$, the model exhibits critical exponents distinct from their short-range counterparts, revealing the influence of long-range interactions while retaining additivity. This condition defines the \emph{weak long-range regime} \cite{mukamel2008statistical, defenu2023longrange}. It is also possible to further subdivide these classes by introducing $\alpha_{\mathrm{MF}}$ such that for $1 < \alpha < \alpha_{\mathrm{MF}}$ the critical exponents correspond to those predicted by mean-field (MF) theories \cite{Dutta2001, Defenu2017}.  It appears that the weak long-range regime is the most challenging to access from an analytical standpoint, making numerical approaches indispensable for its understanding \cite{sun2017, kolokolov1990functional, koziol2021quantumcritical, koffel2012entanglement, kim2023neural}. 
For ease of reference, we summarize this classification in Fig.~\ref{fig:longcla}.

In this paper, we solve the long-range Ising model, which is of significant interest for various experiments \cite{Britton2012, Schau2015, Labuhn2016} and from a theoretical perspective, offering numerous results for comparison. We consider the Hamiltonian given by

\begin{equation}
\label{Hising}
\mathcal H = \sum_{ij}^N J_{ij} \sigma_i^z \sigma_j^z\ - h_x \sum_i^N \sigma^x_i,
\end{equation}
where the $\sigma^{x}$ represent the standard Pauli matrices and $J_{ij}$ is defined as in Eq.~\eqref{eq:Jalpha}. Throughout all computations, we set $h_x = 1$, thereby fixing the units of $J$. This model yields exact analytical results in the limit  $\alpha \to \infty$ (nearest neighbors) via the Jordan-Wigner transformation. Conversely, in the limit $\alpha \to 0$, the model reduces to the Lipkin-Meshkov-Glick (LMG) model, which is also solvable in the thermodynamic limit. Moreover, for $\alpha < d$, the model admits an exact solution in the ferromagnetic ($J<0$) case \cite{romanroche2023}. These analytical insights can be complemented by numerical approaches that span various ranges of $\alpha$, utilizing diverse techniques such as  quantum Monte Carlo \cite{koziol2021quantumcritical, gonzalezlazo2021finitetemperature} and ansätze such as tensor networks \cite{koffel2012entanglement, sun2017}, and, more recently, restricted Boltzmann machines (RBM) \cite{kim2023neural}.
All these  works together  offer a comprehensive overview for both the ferromagnetic and antiferromagnetic scenarios. The system undergoes a second-order quantum phase transition (QPT) from ferromagnetic and Néel respectively ordered phases to paramagnetic phases at a certain critical point $J_c (\alpha)$. Within this model, $\alpha_{MF} = 5/3$, below which the critical exponents align with those predicted by Mean Field theory. Conversely, for $\alpha > 3$, the system falls within the universality class of the 2D classical short-range Ising model. As a consequence, model \eqref{Hising} provides an excellent baseline for evaluating the accuracy of the transformer variational ansatz across the entire $\alpha$ spectrum, ranging from short-range to flat interactions, and for both ferromagnetic and antiferromagnetic cases.

\section{Methods}\label{sec:methods}

\subsection{Variational quantum Monte Carlo}
\begin{figure*}[t!]
    \centering
    \includegraphics[width = \textwidth]{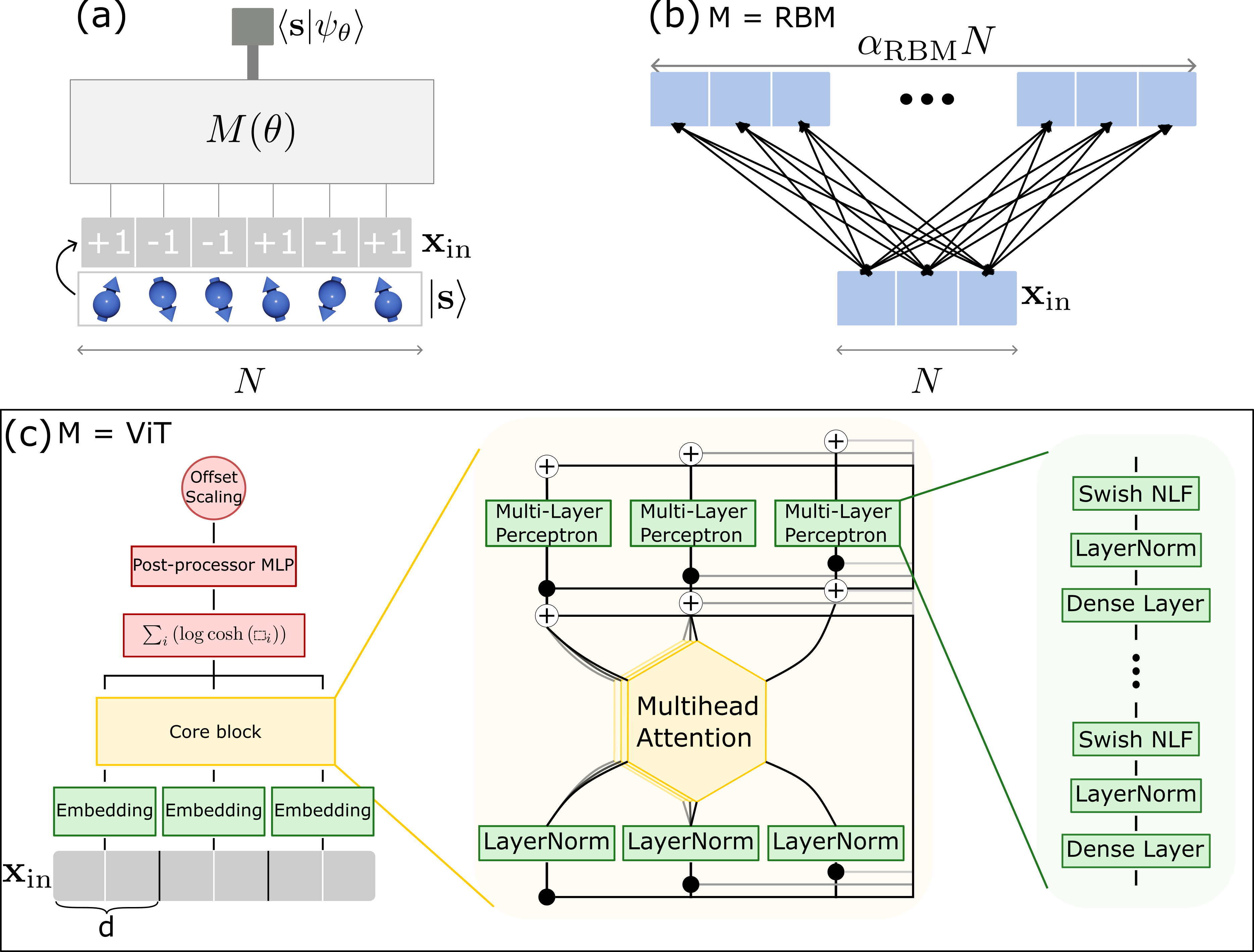}
    \caption{(a) General scheme of the parameterization of a wave-function, $\psi_\theta$, through a machine learning model, $M$, characterized by a series of variational parameters $\boldsymbol{\theta}$. (b) General structure of a RBM. The parameter $\alpha_{\mathrm{RBM}}$ indicates the density of the hidden layer. (c) Block diagram representing the structure of the ViT ansatz used in this work. The color code is used: green corresponds to all those operations that act at token level; in yellow we represent operations that generate correlations between the channels originating from different tokens; finally, red highlights the operations placed after the Core block that merge the channels and post-process that result.}
    \label{fig:ansatze_scheme}
\end{figure*}

We use an unnormalized representation of the ground state over the canonical basis states $\left\lbrace | {\mathbf s} \rangle\right\rbrace$:

\begin{equation}\label{eq:ansatz_wf}
|\psi_{\boldsymbol{\theta}} \rangle = \sum_{{\mathbf s}} \psi_{\boldsymbol{\theta}} (\mathbf{x}_{\mathrm{in}}) | {\mathbf s} \rangle,
\end{equation}

\noindent where $\psi_{\boldsymbol{\theta}}$ is a sufficiently flexible ansatz with a collection of parameters $\boldsymbol{\theta}$, yielding the probability amplitude of each of those states, and $\mathbf{x}_{\mathrm{in}}$ is an appropriate encoding of $| {\mathbf s} \rangle$ to serve as an input to the function. In this work we focus on spin $1/2$ chains, so $\mathbf{x}_{\mathrm{in}}$ will be composed of a binary variable for each spin, whose values we encode as $+1$ or $-1$ (see Figure~\ref{fig:ansatze_scheme}a). Our choices for $\psi_{\boldsymbol{\theta}}$ are all general machine-learning models, which we generically represent as $M\left(\boldsymbol{\theta}\right)$.

In this framework, the solution for the ground state of the system comes in the form of an optimal set of parameters $\boldsymbol{\theta}$. Those parameters are thus variationally optimized by minimizing the energy of the Hamiltonian,
$\langle\mathcal{H}\rangle_{\psi_{\boldsymbol{\theta}}} \equiv \langle \psi_{\boldsymbol{\theta}} \vert \mathcal{H} \vert \psi_{\boldsymbol{\theta}}\rangle / \langle  \psi_{\boldsymbol{\theta}} \vert \psi_{\boldsymbol{\theta}} \rangle$. The expected value of an observable explicitly represented on this basis as a function $Q\left(\mathbf{x}_{\mathrm{in}}\right) = \frac{\langle \mathbf{s} | Q | \psi_{\boldsymbol{\theta}} \rangle}{\langle \mathbf{s} | \psi_{\boldsymbol{\theta}} \rangle}$, can be calculated as

\begin{equation}\label{eq:expec_valueMC}
\langle Q \rangle_{\psi_{\boldsymbol{\theta}}} = \sum_{{\mathbf s}} p_\theta({\mathbf s}) Q\left(\mathbf{x}_{\mathrm{in}}\right)\ ,
\end{equation}
where $p_{\theta}(\mathbf{s}) = |\psi_{\boldsymbol{\theta}}\left(\mathbf{x}_{\mathrm{in}}\right)|^2 / \langle \psi_{\boldsymbol{\theta}}  | \psi_{\boldsymbol{\theta}} \rangle $. Note that, in contrast to a more conventional ML workflow for building a classifier or a regressor, no training outputs are necessary here to define a loss, since optimization is guided by the minimization of the value of the energy $\langle\mathcal{H}\rangle_{\psi_{\boldsymbol{\theta}}}$ estimated using the model itself. The data used to train the model are thus only inputs in the form of states sampled from the canonical basis.

As these definitions suggest, we are forced to resort to Monte Carlo (MC) sampling in order to evaluate the averages to carry out the minimization process, given the impossibility of considering the totality of states. Specifically, we use the Metropolis-Hastings algorithm with a combined transition rule. On the one hand, a random spin is flipped according to a uniform probability without taking into account any constraints such as the total magnetization. On the other hand, we invert the total magnetization of a random number of Markov chains. The probability for the former is three times larger than the probability for the latter in our implementation.
Furthermore, among all the available optimization methods, we employ Stochastic Gradient Descent (SGD) with custom schedules for the learning rate, $\lambda$, combined with the Stochastic Reconfiguration (SR)  method \cite{PhysRevB.61.2599}, characterized by a stabilizing parameter called diagonal shift that we denote as $\Delta_{sr}$.

\subsection{Ansätze: neural networks and restricted Boltzmann machines}\label{subsec:Ansatze}
In this variational framework, NNs are leveraged to provide a trainable $\psi_{\boldsymbol{\theta}}$. In this study we employ and compare two different architectures with very different levels of complexity: an RBM adapted for regression and a vision transformer. Multilayer perceptrons (MLPs), the most basic incarnation of a feed-forward neural network (FFNN), are used as building blocks in the ViT and also serve to introduce this RBM.

The architecture of an FFNN is characterized by the unidirectional flow of information from the input towards the output, passing through a set of hidden layers. In the case of an MLP, each neuron in a layer computes a linear combination of all the outputs from the previous layer and applies a nonlinear activation function to the results to create its own output. The defining characteristics of an MLP are the all-to-all connectivity between adjacent layers and the fact that there is at least a single hidden layer. The functional form of an MLP with activation function $F$ and $\ell$ hidden layers is

\begin{align*}
&\MLP\left(\mathbf{x}_{\mathrm{in}};\mathbf{w}_1,\mathbf{b}_1,\mathbf{w}_2,\mathbf{b}_2\ldots\mathbf{w}_{\mathrm{out}},\mathbf{b}_{\mathrm{out}}\right) = \\
 & = F\left(\mathbf{b}_{\mathrm{out}} + \mathbf{w}_{\mathrm{out}} F\left(\mathbf{b}_{\ell} + \mathbf{w}_{\ell} F\left(\ldots F\left(\mathbf{b}_1 + \mathbf{w}_1 \mathbf{x}_{\mathrm{in}}\right)\right)\right)\right)
\end{align*}

\noindent Here, the $\mathbf{w}$ matrices and $\mathbf{b}$ vectors are the weights and biases of each layer, respectively. Layers can have different widths, and $F$ is understood to act elementwise on its arguments. We create the functional form of our ansatz by using an MLP for the logarithm of the wave function:

\begin{equation}
\label{psi_FF}
\psi_{\boldsymbol{\theta}}^{\text{FFNN}} = \exp\left[\sum\MLP\left(\mathbf{x}_{\mathrm{in}}; \boldsymbol{\theta}\right)\right].
\end{equation}

\noindent where $\boldsymbol{\theta}$ stands for the collection of weights and biases and the sums run over elements of the output to generate a single value. 
This results in a real wave function with positive coefficients, which aligns with our interests. However, if one were to deal with problems where the ground state is a non-positive wave function, the ansatz would need to be modified. One option can be to add a second MLP to parametrize the imaginary part of the logarithm, i.e., the phase of the wave function. This covers both the scenario where the wave function may be complex and the case where it is real but non-positive. For the latter, one can modulate the imaginary part with a sigmoid function with a maximum amplitude of $\pi$ to restrict the phase to the values 0 or $\pi$.

A perceptron with a single hidden layer and some minor changes is equivalent to a restricted Boltzmann machine (RBM; see Figure~\ref{fig:ansatze_scheme}b), originally devised as a generative model. Specifically, the nonlinear activation function must be chosen as $F\left(x\right)=\log\cosh\left(x\right)$ and an additional linear term is added to the output. The ansatz in this case is

\begin{equation}
\begin{aligned}
\log \psi_{\boldsymbol{\theta}}^{\text{RBM}} = \mathbf{a} \mathbf{x}_{\mathrm{in}} +  &\sum_j\log\left[\cosh \left( \left(\Re \mathbf{w}\right) \mathbf{x}_{\mathrm{in}} + \Re \mathbf{b}\right)_j\right] +\\
& i\sum_j\log\left[\cosh \left( \left(\Im \mathbf{w}\right) \mathbf{x}_{\mathrm{in}} + \Im \mathbf{b}\right)_j\right]
\end{aligned}
\end{equation}
In the expression above, the product runs over outputs. The width of the output layer is usually chosen as proportional to the width of the input layer through an integer factor known as the density \cite{Carleo2017}. This variational ansatz results in a wave function with parameters $\boldsymbol{\theta} = \{\mathbf{a}, \mathbf{w}, \mathbf{b}\}$. We take $\mathbf{a} = 0$ following the implementation in \software{NetKet} \cite{netket3:2022}.

Although on paper MLPs can be used to approximate any sufficiently well-behaved function according to a number of universality theorems \cite{cybenko1989approximation, hornik1989multilayer}, the basic MLP architecture faces important limitations in practice arising, among other factors, from the phenomenon of vanishing gradients during training \cite{Goodfellow} as well as the complexity of their all-to-all mapping. The former has been alleviated through breakthroughs such as normalization layers (e.g. BatchNorm \cite{BatchNorm} and LayerNorm \cite{LayerNorm}), residual learning by means of additive shortcuts around groups of layers \cite{ResNet, RegressionResNet} and new activation functions line Swish \cite{swish}. In terms of reduction in complexity, the transformer architecture, originally developed to replace recurrent architectures in machine translation tasks \cite{Vaswani2017}, has found applications in all areas of machine learning, and in particular been very successfully adapted to computer vision tasks as the vision transformer (ViT) \cite{vision_transformer}. A key feature of transformer architectures is their decoupling of nonlinear processing, which is still performed by modified MLPs acting on subsets of the data, and mixing between those pieces of data according to the relevance of those interactions to the result, which is handled by an attention mechanism. The details of the version of the transformer used in this work are presented in the form of a block diagram in Figure~\ref{fig:ansatze_scheme}c, and the main operations are described in what follows.

First of all, \emph{tokenization} takes place. Here the quantum states \( |{\mathbf s}\rangle \), represented as their corresponding arrays $\mathbf{x}_{\mathrm{in}}$, are divided into \( n \)-tokens of dimension \( d \) (\( N = d \, n \)). Next they are passed through a linear embedding layer with trainable parameters (common to all tokens), mapping each token into a vector of arbitrary dimension $d_{\text{emb}}$. The next stage is an attention block where correlations between tokens are taken into account. As depicted in the central panel of Figure~\ref{fig:ansatze_scheme}c, we first normalize the output of the previous embedding using a LayerNorm operation \cite{LayerNorm}. We then split each token into $h$ vectors of dimension $p = d_{\text{emb}}/h$.
Each of these vectors goes into one of the $h$ heads of the multihead attention layer, where the following operation takes place:
\begin{equation}\label{eq:attention}
\mathbf{A}_i^{\mu} = \sum_{j=1}^{n} a_{ij}^{\mu} V^{\mu} \mathbf{x}^\mu_j \; .
\end{equation}
Here, $\mathbf{x}^\mu_j$ represents the $p$-dimensional vector processed by the $\mu$-th head coming from token $j$. As it can be seen, tokens are mixed within each head through the linear transformation parameterized by $V^\mu$ and the attention matrix $a_{ij}^\mu$. Note that this is a simplified version of the original dot-product attention mechanism \cite{Vaswani2017}. The key feature of this simplification is that the attention weights are dependent only on the relative positions of each pair of spins. In contrast, the usual dot-product attention is invariant with respect to permutations and therefore does not encode positions at all, so explicit positional encodings must be introduced to avoid losing that information \cite{Shaw2018}. Furthermore, since we deal with spin chains with periodic boundary conditions, we enforce the translational symmetry by constraining $a^\mu_{ij}$ to be a circulant matrix, i.e., $a^\mu_{ij}=c^\mu_{(j-i) \quad {\rm mod} p}$ \cite{Viteritti2023}. This is complemented by an explicit symmetrization through an average over all the possible cyclic permutations of the tokens. In this regard, the ViT improves upon the complexity of a fully-connected model when applied to a translationally symmetric system, as the latter would require a similar symmetrization over all cyclic permutations of spins. After applying this layer, we concatenate the $h$ $\mathbf{A}_i^{\mu}$ vectors to recover a single $d_{\text{emb}}$-dimensional vector for each of the input tokens. Each of those vectors is then fed through a modified MLP for nonlinear processing, whose components are represented in the right-hand-side panel of Figure~\ref{fig:ansatze_scheme}c. The output is again a $d_{\text{emb}}$-dimensional vector for each input token. Note that both operations (mixing by the attention block and processing by the MLP) are also surrounded by ResNet-like bypasses \cite{RegressionResNet}. We further pass the output through an elementwise $\log\cosh\left(x\right)$, which we have found to greatly enhance the training performance of the model when it must learn wave functions whose norm is concentrated over a few states. The inspiration for our choice of transformation comes from the simpler RBM-like architecture, where the $\logcosh$ function is essential for a satisfactory training, as opposed to more conventional nonlinear activation functions.
This completes our ViT core block and the set of operations defined terms of tokens. A pooling operation is now applied so that the $n$ vectors corresponding to the tokens are averaged to give rise to a single (still $d_{\text{emb}}$-dimensional) vector. This is further processed by another MLP where layer by layer the dimensions of the vector resulting from pooling are reduced: $d_{\text{emb}} \to d_1 \to \dots \to d_{n_{pp}}$. Finally, a last linear transformation acts as an offset-and-scaling block. This complete ViT is used to create the real parts of $\log \psi_{\boldsymbol{\theta}}$, and we do not include an imaginary part.

\subsection{Parameters and computational details}\label{subsec:numdetails}

All calculations in the following studies were performed using GPUs. The hyperparameters used in the ViT architecture are listed in Table \ref{tab:ViThyper}. Our code is written in \software{Python} and available on GitHub \footnote{\url{https://github.com/Theory-and-Simulation-at-INMA/transformer_LR_WF_public/tree/v24.07.1}, doi:10.5281/zenodo.12654460, retrieved on 2024-07-04.}. Our models are defined using the \software{Flax} neural-network library \cite{flax}, which runs on top of \software{JAX} \cite{jax2018github}. Training (including the schedules for the learning rate and for the diagonal shift in the SR preconditioner) is handled by a combination of \software{Optax} \cite{deepmind2020jax} and \software{NetKet} \cite{netket3:2022}, which also takes care of sampling.

\begin{table}[h!]
\centering
\resizebox{\columnwidth}{!}{%
\begin{tabular}{ccc}
\toprule
\textbf{Variable} & \textbf{Notation} & \textbf{Value} \\ 
\midrule
Chain length & $N$ & $50-200$ \\ 
Token dimension & $d$ & $N/10$ \\ 
Number of tokens & $n = N/d$ & $10$ \\ 
Embedding dimension & $d_{\rm{emb}}$ & $14$ \\ 
Number of heads & $h$ & $2$ \\ 
Head dimension & $p = d_{\rm{emb}}/h$ & $7$ \\ 
MLP layers within Core Block & $n_{\text{MLP}}$ & $3$ \\ 
Post-processor MLP & $(d_1,\dots ,d_{n_{pp}})$ & $(5,)$ \\ 
\bottomrule
\end{tabular}%
}
\caption{Summary of parameters in the ViT architecture and their values used for the simulations [cf. Figure \ref{fig:ansatze_scheme}].}
\label{tab:ViThyper}
\end{table}

Each iteration of the training is performed using a total of $4096$ MC samples distributed over $1024$ independent Markov chains. For the training process we chose a protocol for the learning rate, $\lambda$, consisting of a linear warm-up followed by an exponential decay. This protocol is defined by an initial value of the learning rate, $\lambda_0$, a maximum value, $\lambda_{\mathrm{max}}$ that it attains after $n_{\mathrm{warm}}$ iterations corresponding to the linear warm-up and a ratio $\gamma$ for the exponential decay. To obtain the states with which we created Figure \ref{fig:PhaseDiagram}, a maximum of $250$ iterations for training were used, along with following parameters: $\lambda_0 = 0.1$, $\lambda_{\mathrm{max}} = 2.0$, $n_{\mathrm{warm}} = 75$, $\gamma = 0.995$. The stabiliser parameter present in the SR method follows a simple linear schedule with an initial value of $\Delta_{sr} = \num{e-2}$ and a final value of $\Delta_{sr} = \num{e-4}$.

\section{Results}\label{sec:results}

Here we discuss the performance of the ViT ansatz for characterizing the full phase diagram of Eq.~\eqref{Hising}. We compare these results with previous studies that cover specific ranges of the model.

\subsection{Quantities to compute}
We define the \emph{generalized} staggered magnetization through the operator $m_s$,
\begin{equation}\label{eq:staggered_mag}
    \hat m_s\left(q\right) = \frac{1}{N}\sum_j \hat\sigma^z_j e^{iqj}\ ,
\end{equation}
where $q = 0$ when $J < 0$ and $q = \pi$ when $J > 0$. 
Once the network parameters have been determined, this can be computed using Eq.~\eqref{eq:expec_valueMC}.

To characterize the entanglement between two partitions of the system within the ground state, we also compute the Renyi-2 entanglement entropy $\hat S_2$ following Ref.~\citenum{Hastings2010}. We define $\vert\psi_{\boldsymbol{\theta}}\rangle$ based on the states of the computational basis corresponding to each of the partitions, $\mathbf{s}_A\in A$ and $\mathbf{s}_{\bar{A}}\in\bar{A}$, such that
$\vert\psi_{\boldsymbol{\theta}}\rangle = \sum_{\mathbf{s}_A, \mathbf{s}_{\bar{A}}} \psi_{\boldsymbol{\theta}}(\mathbf{s}_A, \mathbf{s}_{\bar{A}})\vert \mathbf{s}_A\rangle \vert\mathbf{s}_{\bar{A}}\rangle$. Therefore we can define
\begin{equation}\label{eq:renyi2}
    \hat S_2 = -\log_2\Tr[\rho_A^2]\ ,
\end{equation}
where $\rho_A = Tr_{\bar{A}}\rho$.
%
%
From this definition we can sample states from each of the partitions to estimate $\langle \hat S_2\rangle$ according to
\begin{equation}\label{eq:S2_MC}
    \langle \hat S_2\rangle = \sum_{\substack{\mathbf{s}_A, \mathbf{s}_{\bar{A}} \\
    \mathbf{s}'_A, \mathbf{s}'_{\bar{A}}}
    } p_\theta(\mathbf{s}_A, \mathbf{s}_{\bar{A}})p_\theta(\mathbf{s}'_A, \mathbf{s}'_{\bar{A}}) S_2(\mathbf{s}_A, \mathbf{s}_{\bar{A}}, \mathbf{s}'_A, \mathbf{s}'_{\bar{A}})\ ,
\end{equation}
where
\begin{equation}\label{eq:S2_sampling}
    S_2(\mathbf{s}_A, \mathbf{s}_{\bar{A}}, \mathbf{s}'_A, \mathbf{s}'_{\bar{A}}) = -\log \left\langle \frac{\psi_{\boldsymbol{\theta}}(\mathbf{s}_A, \mathbf{s}'_{\bar{A}}) \psi_{\boldsymbol{\theta}}(\mathbf{s}'_A, \mathbf{s}_{\bar{A}})}{\psi_{\boldsymbol{\theta}}(\mathbf{s}_A, \mathbf{s}_{\bar{A}}) \psi_{\boldsymbol{\theta}}(\mathbf{s}'_A, \mathbf{s}'_{\bar{A}})} \right\rangle \ .
\end{equation}
For our simulations we choose $A$ to be half of the system size. This procedure is already implemented in NetKet \cite{netket3:2022}.

Finally, it is useful to employ a metric to quantify the quality of the converged wave functions.  One such metric is the V-score \cite{wu2023variational}. It is defined in terms of the energy fluctuations, conveniently normalized to yield an \(N\)-independent quantity:
\begin{equation}\label{eq:vscore}
\mathrm{V\mhyphen score} = N \left( \frac{ \langle \mathcal{H}^2 \rangle_{\psi_{\boldsymbol{\theta}}} - \langle \mathcal{H} \rangle_{\psi_{\boldsymbol{\theta}}}^2 }{\langle \mathcal{H} \rangle_{\psi_{\boldsymbol{\theta}}}^2}\right)\ ,
\end{equation}

\subsection{Phase diagram and critical exponents}

\begin{figure*}[t!]
\centering
\includegraphics[width=\textwidth]{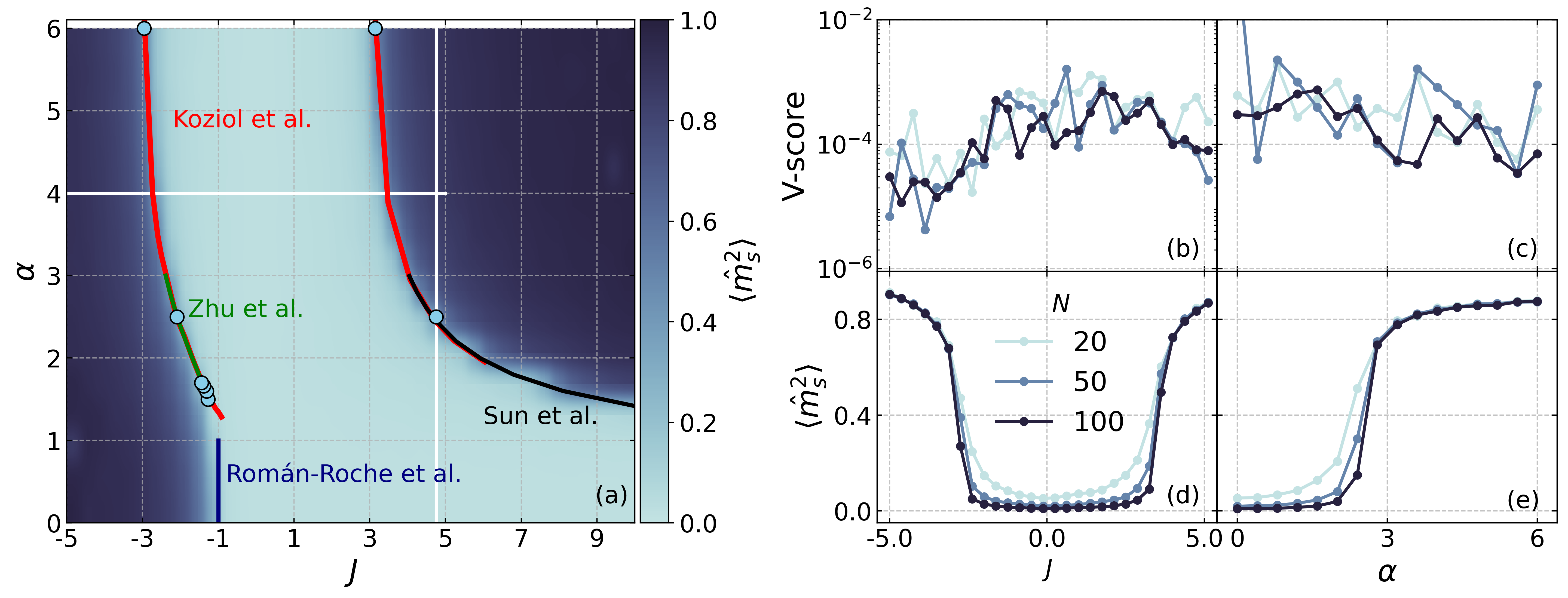}
\caption{(a) Squared magnetization, $\langle\hat m^2_s\rangle$, obtained from the ground states determined by the ViT ansatz for a size of $N = 50$ spins for different values of $J$ and $\alpha$. In this diagram different transition curves present in the literature are represented \cite{koziol2021quantumcritical, zhu2018fidelity, sun2017, romanroche2023}. In addition, the points indicated with light blue circles correspond to our estimates obtained by finite-size scaling analysis. The panels to the right of the phase diagram indicate the behavior of different quantities along two slices of the diagram (white lines). In panels (b) and (d) the value of $\alpha = 4.0$ has been set and the V-score \eqref{eq:vscore} and squared magnetization $\langle\hat m^2\rangle$ are shown as a function of $J$, respectively. In panels (c) and (e) the same quantities are shown in the same order but fixing $J = 4.75$ and varying $\alpha$. The different curves in all these panels indicate different chain sizes, with $N = 20$ being the lightest color and $N = 100$ the darkest, going through $N =50$ (intermediate).}
 \label{fig:PhaseDiagram}
 \end{figure*}

In Figure \ref{fig:PhaseDiagram} we summarize our calculations for the full phase diagram  of the Ising model \eqref{Hising}. In contour \ref{fig:PhaseDiagram}a) we plot the fluctuations $\langle \hat m_s ^2 \rangle$ in order to resolve the phase transition.  Here the simulations are done with $N = 50$ spins and a discretization of $\delta J = 0.375$ and $\delta \alpha = 0.4$.  The darker zones stand for ordered phases, both ferro and antiferromagnetic.   We also plot several critical $J_c$ found with different techniques in previous works \cite{koziol2021quantumcritical, zhu2018fidelity, romanroche2023, sun2017}. Light blue circles mark values where a deeper investigation has been done (see below).
In panels \ref{fig:PhaseDiagram} b) and c), the V-score is plotted for different sizes \(N=20, 50, 100\)  along the cuts marked in panel a) with white lines (the darker the curve, the larger the size). We observe that the V-score does not depend on the system size, demonstrating that the simulations do not deteriorate as \(N\) increases. The values around \(10^{-4}\) compare well with variational methods for other quantum many-body problems \cite{wu2023variational}. Beyond this figure of merit, in panels \ref{fig:PhaseDiagram} d) and e) we plot the fluctuations \(\langle \hat{m}_s^2 \rangle\) for the same sizes and along the same cuts. In addition to the finite-size  behavior, the curves are smooth, which confirms the convergence of the method.  It is important to note that all of these results are the outcome of single shot runs. No averaging or filtering has been performed, which again highlights the consistent convergence of the model. It is worth stressing that the inclusion of the $\logcosh$ function at the end of the core block [cf. Figure \ref{fig:ansatze_scheme}], as described in Section \ref{subsec:Ansatze}, is instrumental in obtaining accurate states in the deep FM regime.

Given that the model is able to qualitatively obtain the phase diagram across the entire parameter space, we proceed to extract the critical exponents. This requires precision in the order parameter that characterizes the transition. In our case, we take advantage of how well the ViT reproduces the fluctuations \(\langle \hat{m}_s^2 \rangle\). This allows us to determine both the critical point \(J_c\) and the critical exponents associated with the phase transition, \(\nu\) and \(\beta\).
In order to do so, we use 
finite size scaling theory \cite{newman1999monte, binder2012monte}. Close the critical point, $\langle \hat m_s^2\rangle$  scales with the system size, $N$, as
\begin{equation}
    \langle \hat m_s^2(N, J)\rangle = N^{-2\beta/\nu} f\left(N^{1/\nu} (J - J_c)\right) \ ,
\end{equation}
where $\nu$ is the critical exponent indicating how the correlation length diverges and $\beta$ the exponent indicating how the magnetization diverges in the thermodynamic limit. $f$ is a dimensionless scaling function that depends on the ratio between system size and correlation length and controls for finite size effects. The critical values are fitted so that different sizes collapse with the same $f$.   In our case, we use  the Python \software{fssa} library \cite{fssa, melchert2009autoscalepy} for the fit.  All the collapses obtained are plotted in Appendix~\ref{app:FSSA}.
Table \ref{tab:criticalexponents_short} summarizes the critical values obtained for the most paradigmatic cases, comparing them to analytical results, where available, and to other numerical values reported in 
Refs. \cite{koziol2021quantumcritical} and \cite{kim2023neural}.  Note that, to enable a direct comparison, we do not apply the Kac renormalization factor of Eq.~\eqref{eq:Kac_factor}. 
In addition to $J_c$, $\nu$ and $\beta$, we also list  $\tilde h_c = 1/\tilde J_c$ and  $\theta_c = \arctan(1/\tilde h_c)$. 

\begin{table}[h!]
\centering
\resizebox{\columnwidth}{!}{%
\begin{tabular}{ccccccc}
\toprule
\multicolumn{7}{c}{FM} \\
\midrule
$\alpha$ & Method & $J_c$ & $\tilde{h}_c$ & $\theta_c$ & $\nu$ & $\beta$ \\
\midrule
\multirow{3}{*}{1.5} & Ours & $-1.27(3)$ & $4.57(9)$ & $0.22(9)$ & $1.6(5)$ & $0.46(7)$ \\
 & Ref.~\citenum{koziol2021quantumcritical} & $-1.2107$ & $4.7600$ & $0.2071$ & $1.9911$ & $0.4968$ \\
 & Theory & - & - & - & $2.0$ & $0.5$ \\
\midrule
\multirow{2}{*}{2.5} & Ours & $-2.09(1)$ & $1.76(1)$ & $0.516(9)$ & $1.08(9)$ & $0.19(3)$ \\
 & Ref.~\citenum{koziol2021quantumcritical} & $-2.0878$ & $1.7631$ & $0.5159$ & $1.1084$ & $0.1880$ \\
\midrule
\multirow{3}{*}{6.0} & Ours & $-2.963(1)$ & $1.0242(4)$ & $0.7734(2)$ & $1.00(3)$ & $0.12(3)$ \\
 & Ref.~\citenum{koziol2021quantumcritical} & $-2.9444$ & $1.0307$ & $0.7703$ & $0.9849$ & $0.1238$ \\
 & Theory & - & - & - & $1.0$ & $0.125$ \\
\midrule
\multicolumn{7}{c}{AFM} \\
\midrule
$\alpha$ & Method & $J_c$ & $\tilde{h}_c$ & $\theta_c$ & $\nu$ & $\beta$ \\
\midrule
\multirow{3}{*}{2.5} & Ours & $4.7527(6)$ & $0.77449(9)$ & $0.91180(3)$ & $1.17(2)$ & $0.0911(9)$ \\
 & Ref.~\citenum{koziol2021quantumcritical} & $4.6638$ & $0.7893$ & $0.9026$ & $0.9103$ & $0.1153$ \\
 & Ref.~\citenum{kim2023neural} & $4.6778$ & $0.7869$ & $0.9041$ & $1.01$ & $0.122$ \\
\midrule
\multirow{2}{*}{6.0} & Ours & $3.143(2)$ & $0.9655(5)$ & $0.8029(2)$ & $1.04(2)$ & $0.11(2)$ \\
 & Theory & - & - & - & $1.0$ & $0.125$ \\
\bottomrule
\end{tabular}%
}
\caption{List with all critical points and exponents obtained with a ViT ansatz through finite-size scaling analysis (marked as Ours in the Method column) for different values of $\alpha$ and in both interaction regimes, FM ($J_c < 0$) and AFM ($J_c > 0$). Moreover, we show the analytical results when they exist, at $\alpha \leq \alpha_{MF}=5/3$ and $\alpha = 6 > \alpha^*$, as well as different values reported in Refs.~\citenum{koziol2021quantumcritical} and \citenum{kim2023neural}.}
\label{tab:criticalexponents_short}
\end{table}

The values presented in Table \ref{tab:criticalexponents_short} are comparable to those found in the literature or indicated by the theory. During our investigation, we observed that the results worsen for the smallest \(\alpha\), which is unexpected since we are approaching the MF regime where the wave function should be easier to find. Therefore, we attribute this to finite-size effects. This was confirmed by simulating longer chains (\(N=200\)) and observing an improvement in the results.

\begin{figure}[h!]
    \centering
    \includegraphics[width = \columnwidth]{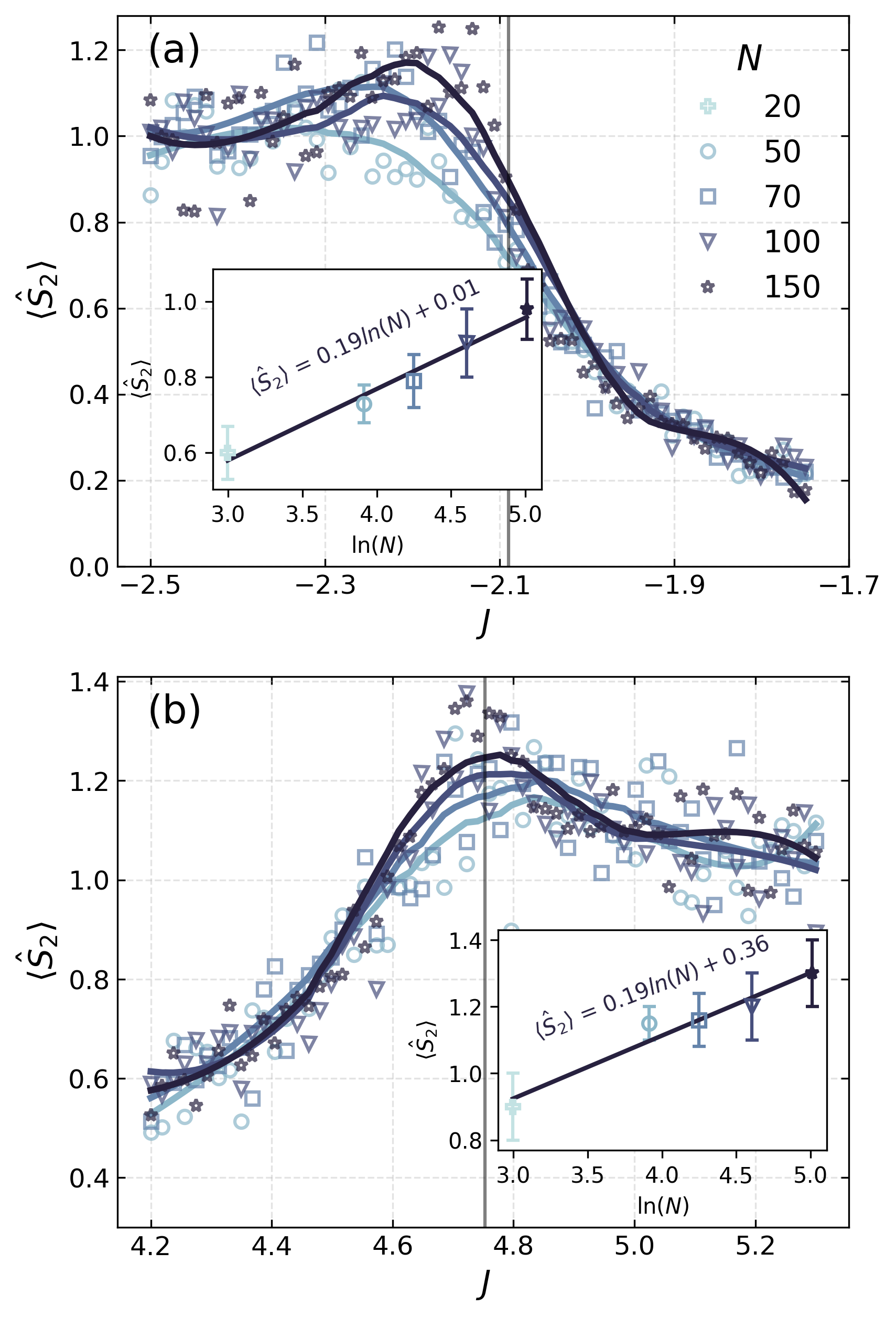}
    \caption{Determination of the Renyi-2 entropy. For $\alpha = 2.5$, the entropy for different chain sizes is shown in the FM regime [panel (a)] and in the AFM regime [panel (b)]. The dots in the main panels represent the simulation results [cf. Eq.~\eqref{eq:S2_MC}] while the solid lines have been obtained with a Savitzky-Golay filter to act as a guide to the eye. The insets show the statistical analysis performed at the critical points of each case (vertical lines in the main panels). The mean value obtained from 10 independent runs is displayed, along with the corresponding standard deviation as error bars. The linear fit indicates that the system's entropy follows the expected logarithmic scale in both cases.}
    \label{fig:entropy_flucts}
\end{figure}

Once the critical point has been determined, we can study the behavior of the entanglement entropy. On one hand, it reveals the limitations in the amount of entanglement other methods such as MPS can capture. On the other hand, it also characterizes the critical points, where it is known that the behavior of the entropy is universal and violates the area law, following a logarithmic dependence with the size $N$ \cite{Calabrese2009}. Our results for the entropy are plotted in Figure \ref{fig:entropy_flucts} as a function of \(J\) for both ferromagnetic and antiferromagnetic cases at \(\alpha = 2.5\).
These points correspond to single-shot executions. To aid visualization, we applied a Savitzky-Golay filter, which allows us to observe a certain trend across the different system sizes. Far from the critical point, indicated by a solid gray line in each case corresponding to the values reported in Table \ref{tab:criticalexponents_short}, it is clear that the entropy becomes size-independent, in accordance with the area law. However, near the critical point, these data do not allow us to deduce the predicted volume dependence. Therefore, we focus on the estimated critical point and perform statistics over 10 independent runs. The insets in each of the panels of Figure \ref{fig:entropy_flucts} show the results for the mean and standard deviation. In this case, it is clear that the data are consistent with the expected logarithmic growth, validating the ansatz’s ability to reproduce highly entangled states.

\subsection{Comparison to the RBM-like solution}

\begin{figure}[h!]
    \centering
    \includegraphics[width=\columnwidth]
    {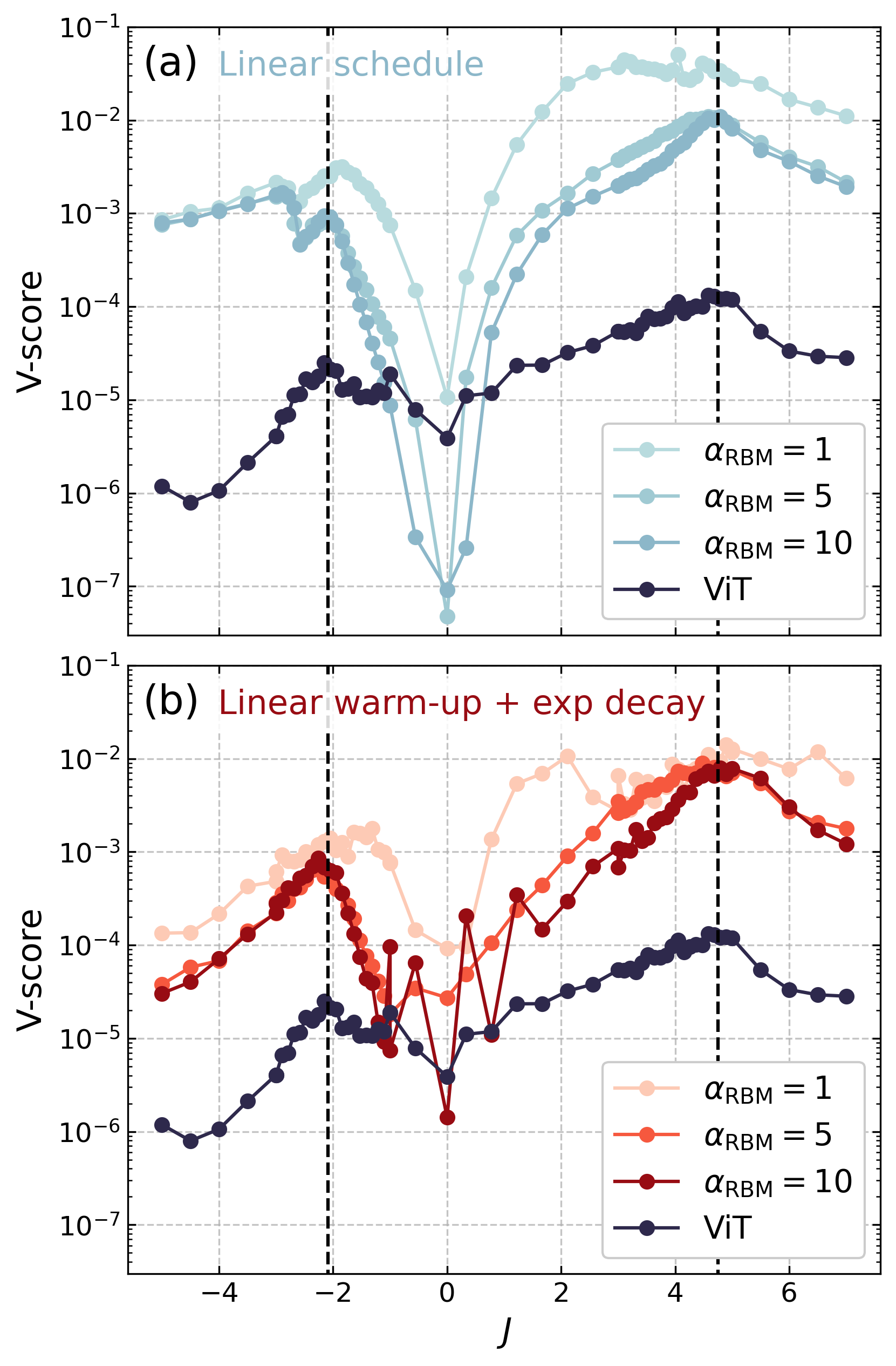}
    \caption{Performance comparison between RBM-like models and ViT. Two different schedules for the learning rate have been used for the RBM, indicated in panels (a) and (b). The training for the ViT model is the same in both cases and it is described in Sec. \ref{subsec:numdetails}.  Fixing a maximum training time (3 minutes) for each point, the V-scores obtained by an RBM-like ansatz with different densities (denoted as $\alpha_{\mathrm{RBM}}$ in the legend) are compared with the ones obtained by our ViT implementation. The chain size is $N = 50$ and the interaction range $\alpha = 2.5$. The black dashed lines mark the critical points obtained by FSSA [see Table \ref{tab:criticalexponents_short}].}
    \label{fig:comparation_vit_rbm}
\end{figure}

We conclude this section by comparing the ViT and RBM-like architectures using the V-score \eqref{eq:vscore} and limiting the total training time for each network to 3 minutes. The results for \(\alpha = 2.5\) and \(N = 50\) as a function of \(J\) are shown in Figure \ref{fig:comparation_vit_rbm}. The training used for the ViT is the same as the one described in Sec.~\ref{subsec:numdetails}. For the RBM, we use hand-optimized parameters for all values of $J$.
However, since the chosen training protocol may be a crucial factor in the model's performance, we explore two different protocols to ensure fairness in the comparison. These two protocols consist in different schedules for the learning rate $\lambda$: a linear decrease and a linear warm-up followed by an exponential decay. The first one corresponds to a linear decrease where the initial and final values of the learning rate are $\lambda_0 = 0.1$ and $\lambda_{\mathrm{min}} = 0.01$, respectively. The other has the same structure as the one detailed in Sec.~\ref{subsec:numdetails}, with values for those parameters of $\lambda_0 = 0.01$, $\lambda_{\mathrm{max}} = 1.2$, $n_{\mathrm{warm}} = 50$ and $\gamma = 0.995$.
The corresponding results for each protocol are shown in panels (a) and (b) of Figure \ref{fig:comparation_vit_rbm}, respectively. To allow for more iterations within the maximum allotted time, the number of MC samples has been reduced to 2048 in all cases.

The main observation from Figure \ref{fig:comparation_vit_rbm} is that the ViT architecture achieves better quantum states across most of the spectrum in all scenarios, including within the critical regions, which are marked by dashed lines. In this regard, it is particularly interesting how the V-score worsens in the vicinity of the critical point for all the architectures and training protocols considered. This behaviour is similar to what has been observed in other ansätze used to obtain ground states such as quantum circuits \cite{roca2023circuit}. Additionally, we see that increasing the RBM's density briefly improves the results, but ultimately reaches a plateau in the V-score that cannot be surpassed by simply adding more hidden neurons to the RBM.
The only region where all architectures converge in accuracy is around $J = 0$, where the model is trivial. In fact, numerical tests indicate that in this limit it is favourable to reduce the token dimension to a single spin, replicating the non-interacting spin scenario, which in turn helps us understand why the simpler RBM achieves better accuracy here. Nonetheless, the ViT provides consistent quality results regardless of the value of $J$ and the phase, whether ordered or not. Moreover, the RBM with \(\alpha_{\text{RBM}} = 1\) already has more parameters than our ViT architecture, with $5100$ vs. $1133$, respectively. Therefore, the ViT is able to achieve better results using fewer parameters and less training time, demonstrating its superiority in terms of efficiency.

Despite these results for the finite-size case, it is worth noting that a recent study \cite{Trigueros2024} demonstrates that the RBM is capable of finding the ground state using only a single variational parameter in the thermodynamic limit ($N\to\infty$) and in the strong long-range regime ($\alpha < 1$).

\section{Conclusions}\label{sec:conclusions}

In this work, we explored the application of the Vision Transformer (ViT) to quantum long-range models, with a focus on the transverse-field Ising model. Our results affirm the ViT's capability to handle all interaction regimes, from short-range to strong long-range interactions. 
Figure \ref{fig:PhaseDiagram} summarizes the full phase diagram and its comparison with previous results that focus on specific regions. Additionally, Table \ref{tab:criticalexponents_short} quantitatively compares the critical properties, $J_c$, and the critical exponents $\nu$ and $\beta$ with values reported in the literature.

Throughout this work, our intention has been to employ an economical approach in terms of the number of hyperparameters and to use single-shot results, without further curation or cherry-picking. The attention mechanism, a fundamental aspect of this architectural design, is responsible for identifying the most significant correlations within a relatively small number of iterations. Additionally, the remaining MLP layers endow the model with sufficient expressiveness and versatility, allowing it to produce a solution for any point within the parameter space. This is further corroborated by Figure \ref{fig:comparation_vit_rbm}, where a comparison was made with an ansatz based on RBMs setting the maximum training wall time. The ViT outperforms the accuracy obtained in critical regions by up to two orders of magnitude. Moreover, the use of the $\logcosh$ function as a nonlinear activation function at the end of the core block [cf. Figure \ref{fig:ansatze_scheme}] was revealed as crucial: its inclusion, as described in Section \ref{subsec:Ansatze}, was fundamental in obtaining consistently and stably accurate states in the deep FM regime.

Our results illustrate the power of modern NN architectures to afford superior accuracy and efficiency than their more traditional counterparts when used as ansätze in variational quantum Monte Carlo workflows, and specifically demonstrate the ability of the ViT to capture important correlations in a cost-effective manner, opening the door to its use for a wider variety of problems involving long-range interactions.

\section*{Acknowledgements}

The authors acknowledge funding through grants CEX2023-001286-S, from 
MCIN/AEI/10.13039/501100011033 and the EU ``NextGenerationEU''/PRTR, and TED2021-131447B-C21,
from MCIN/AEI/10.13039/501100011033.  This study was also supported by MCIN with funding from European Union NextGenerationEU (PRTR-C17.I1) promoted by the Government of Aragon. We also acknowledge the Gobierno de Arag\'on (Grant E09-17R Q-MAD), Quantum Spain and the CSIC Quantum Technologies Platform PTI-001. This research project was made possible through the access granted by the Galician Supercomputing Center (CESGA) to its supercomputing infrastructure. The supercomputer FinisTerrae III and its permanent data storage system have been funded by the Spanish Ministry of Science and Innovation, the Galician Government and the European Regional Development Fund (ERDF). S. R-J.  acknowledges financial support from Gobierno de Arag\'on through a doctoral fellowship.

\appendix

\section{Finite-size scaling analysis}\label{app:FSSA}

\begin{figure*}[h!]
    \centering
    \includegraphics[width=0.9\textwidth]{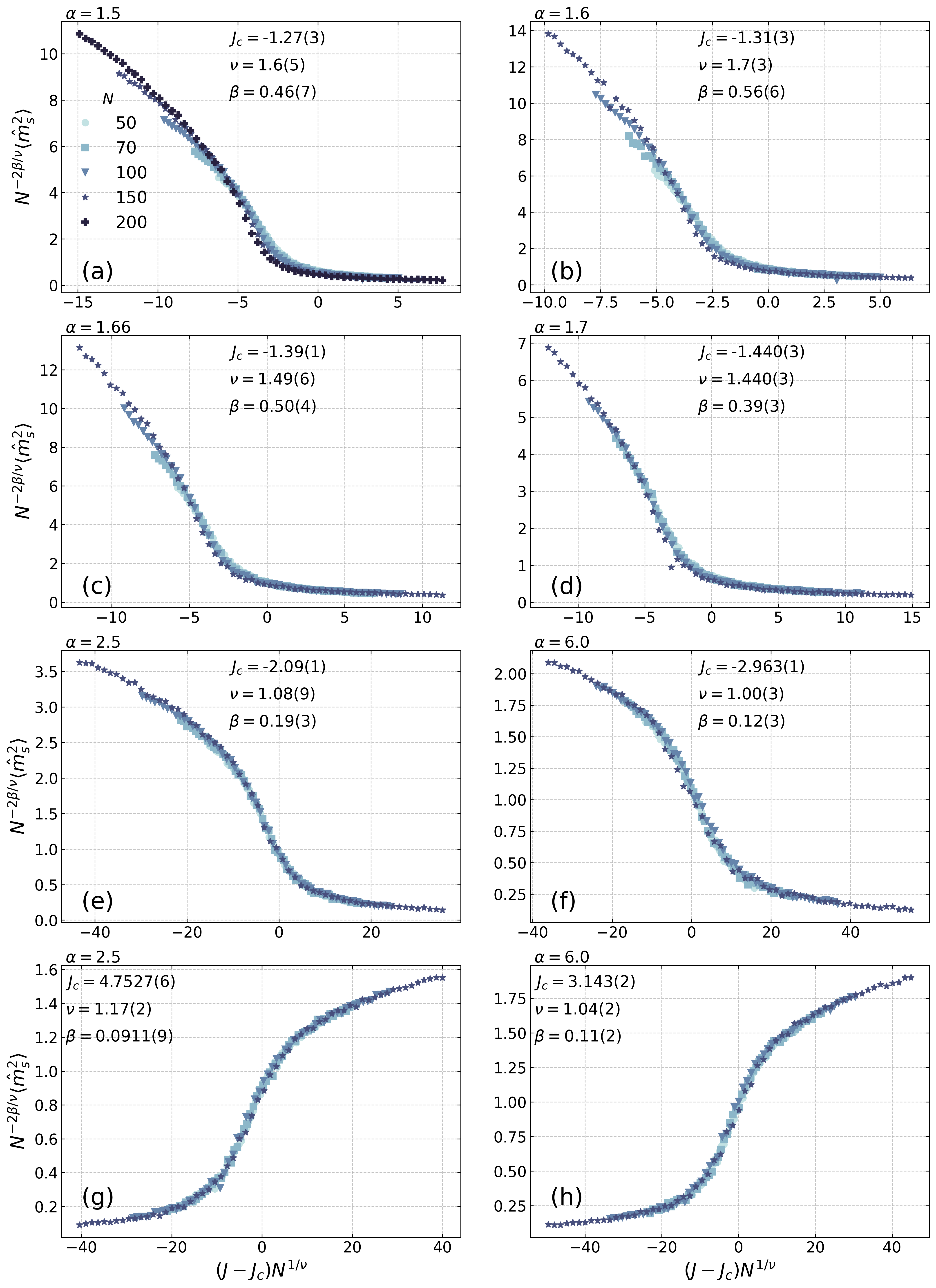}
    \caption{Collapsed curves using the FSSA technique for the data obtained from the fluctuations $\langle \hat m_s^2(N, J)\rangle$ for different values of $\alpha$. In each panel the values of the critical parameters obtained through the corresponding fit are shown.}
    \label{fig:FSSAcurves}
\end{figure*}

Here we show all the collapses of the curves corresponding to the fluctuations that are used to perform finite-size scaling analysis (FSSA), as detailed in the main text, in order to obtain the critical parameters of the system. These curves are shown in Figure \ref{fig:FSSAcurves}. For each value of $\alpha$ considered, $60$ points have been simulated in different windows for $J$. Not every point has been used to obtain the critical parameters; narrower windows centred on the critical point and different for each $\alpha$ have been chosen in order to obtain the best collapses. However, in Figure \ref{fig:FSSAcurves} all the $60$ points are shown. To get these points a maximum number of $500$ iterations was set in the training procedure. The schedule for the SR parameter, $\Delta_{\mathrm{sr}}$, the number of warm-up iterations, $n_{\mathrm{warm}}$, as well as the decay rate $\gamma$, are all the same as the ones described in Section \ref{subsec:numdetails}. The peak value for the learning rate, $\lambda_{\mathrm{max}}$ goes from $5.0$ for the training of the smallest chains ($N = 50$) up to $12.0$ for the largest ($N = 200$) taking intermediate values for the intermediate sizes.

The maximum chain size was $N = 150$ spins for all interaction ranges except for $\alpha = 1.5$, where we made a further push to simulate $N = 200$ spins to verify the finite size effects in the MF regime mentioned in the main text.

\bibliography{main.bib}

\end{document}